\documentclass[a4paper]{PoS}
\usepackage{subfigure}

\title{AMON Searches for Jointly-Emitting Neutrino + Gamma-Ray Transients}

\ShortTitle{AMON Searches for Jointly-Emitting Neutrino + Gamma-Ray Transients}

\author{\speaker{A. Keivani}\thanks{The AMON team URL: http://amon.gravity.psu.edu/participants.shtml}\\
        Department of Physics, Pennsylvania State University,
  University Park, PA 16802, USA\\
        E-mail: \email{keivani@psu.edu}}

\author{D. B. Fox\\
Department of Astronomy \& Astrophysics, Pennsylvania
  State University, University Park, PA 16802, USA\\
          E-mail: \email{dfox@astro.psu.edu}}
          
\author{G. Te\v{s}i\'{c}\\
Department of physics, Pennsylvania
  State University, University Park, PA 16802, USA\\
          E-mail: \email{gut10@psu.edu}}

\author{D. F. Cowen\\
Department of physics, Pennsylvania
  State University, University Park, PA 16802, USA\\
          E-mail: \email{cowen@phys.psu.edu}}
          
\author{J. Fixelle\\
Department of Physics \& Astronomy, Northwestern University, Evanston, IL 60208, USA\\
          E-mail: \email{JoshuaFixelle2014@u.northwestern.edu}}

\abstract{We present the results of archival coincidence analyses using public neutrino data from the 40-string configuration of IceCube (IC40) and contemporaneous public gamma-ray data from Fermi LAT. Our analyses have the potential to discover statistically significant coincidences between high-energy neutrino and gamma-ray signals, and hence, possible jointly-emitting neutrino/gamma-ray transients. This work is an example of more general multimessenger studies that the Astrophysical Multimessenger Observatory Network (AMON) aims to perform. AMON is currently under development and will link multiple running and future high-energy neutrino, cosmic ray and follow-up observatories as well as gravitational wave facilities. This single network will enable near real-time coincidence searches for multimessenger astrophysical transients and their electromagnetic counterparts. We will present the component high-energy neutrino and gamma-ray datasets, the statistical approaches that we used, and the results of analyses of the IC40+LAT datasets.}

\FullConference{The 34th International Cosmic Ray Conference,\\
		30 July- 6 August, 2015\\
		The Hague, The Netherlands}

\begin{document}

\section{Introduction}

The Astrophysical Multimessenger Observatory Network (AMON) links existing and forthcoming high-energy astrophysical observatories into a single virtual system, capable of sifting through the various data streams in near real-time, identifying candidate and high-significance multimessenger transient events, and providing alerts to interested observers~\cite{AMON2013,gociazi}.
The AMON framework also enables searches for coincidences in archival data. This helps us to better comprehend the datasets and also to explore different statistical approaches to generate realtime AMON alerts for the network's followup partners. 

Many theoretical models predict that detectable neutrino emission will be paired with a prompt electromagnetic signal~\cite{wb,rachen}.
Recently, the IceCube Observatory reported an excess of astrophysical neutrinos in the $10^{13-15}$ eV energy scale~\cite{IC3yrs}. 
These motivate us to perform archival coincidence analyses between the IceCube public data and the Fermi LAT public photon datasets. Our goal is to look for statistically significant coincidences between high-energy neutrinos and gamma-ray signals to find jointly-emitting neutrino/gamma-ray transient sources. 

This paper is organized as follows: The detail of the datasets we use is discussed in section~\ref{data}. Different statistical methods are discussed in section~\ref{stat}. Section~\ref{results} shows the results of the analyses. Conclusions and future work are presented in section~\ref{conclusions}.

\section{Datasets}
\label{data}

This analysis has been performed using IceCube and Fermi LAT public data on a period of temporal overlap between the two observatories. IceCube public data includes the 40-string and 59-string configurations of IceCube observatory (IC40 and IC59). IC40 dataset starts on April 6, 2008 and ends on May 20, 2009 and IC59 dataset is between May 20, 2009 and May 31, 2010. The public Fermi LAT dataset starts on August 4, 2008 and has been continuously available since then. The overlap period of IC40 and Fermi LAT data is therefore between weeks 9 and 50 of the Fermi LAT data. This period for IC59 is between weeks 50 and 104. 

The IC40 public data only contains neutrinos from the northern hemisphere (positive declination). In total, there are about 14,000 neutrinos in this dataset. For the purpose of this analysis, we only consider the northern hemisphere data of IC59 as well which contains about 43,000 neutrinos. 
The Fermi LAT data that is being used is selected by imposing additional criteria using Fermi LAT purest (i.e. lowest instrumental background) analysis class (Pass 7-V6 Ultraclean). In this class, the photon events that are detected while the telescope is repositioning, or is in pointing mode, or pointing to the Sun are removed. In addition, only photon events with spacecraft zenith angle smaller than 65$^\circ$ and energies above 200 MeV are accepted. 
%http://fermi.gsfc.nasa.gov/ssc/data/analysis/LAT_essentials.html
%\begin{itemize}
%\item
%Telescope repositioning: the photons detected while the telescope is changing its pointing direction are removed.
%\item
%Telescope pointing: the photons detected while the telescope is placed in pointing mode (i.e. conducting a steady state source observation) are removed.
%\item
%Solar pointing: the photons detected while the telescope is looking at the Sun are removed.
%\item
%Energy: any photon with energy below 200 MeV is removed as the effective area falls off rapidly below this energy.
%\item
%Photon-spacecraft zenith angle: any photon detected beyond the spacecraft zenith angle of 65$^\circ$ is removed.
%\end{itemize}

\begin{figure}
\begin{center}
\subfigure[Fermi LAT]{
\includegraphics [trim=0cm 0cm 0cm 0cm,scale=0.47]{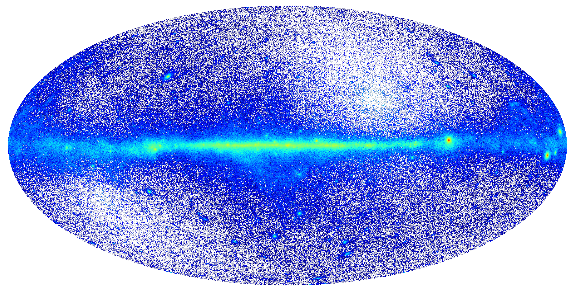}}
\subfigure[IC40+IC59]{
\includegraphics [trim=0cm 0cm 0cm 0cm,scale=0.35]{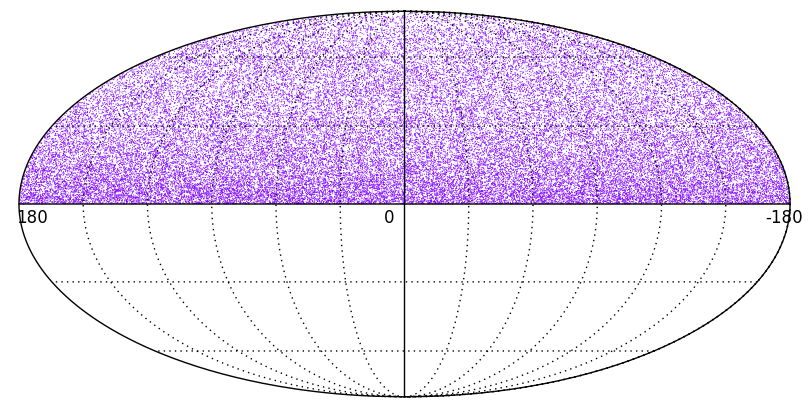}}
\caption{Sky map in equatorial coordinates for (a) Fermi LAT photons on a period of temporal overlap with IC40+IC59 and (b) IC40+IC59 neutrinos.}
\label{skymap}
\end{center}
\end{figure}

Applying all these criteria, the dataset reduces to approximately 4.1 million photon events for the period overlapping IC40 and 5.5 million photon events for the period overlapping IC59. 
Figure~\ref{skymap} shows the Fermi LAT exposure corrected map for the IC40+IC59 period (on the left) and the observed neutrinos in northern hemisphere of IC40+IC59 data (on the right). 
There is an excess flux above the horizon in the neutrino map that is presumably the result of cosmic ray muons that are incorrectly reconstructed as coming from the northern sky. 
Both maps are in equatorial coordinates. 

To start the analysis, we calculate the angular separation between each photon event and each neutrino and identify all pairs with smaller than 10$^\circ$ separation and $|\Delta\rm{t}|<50$ s. This value of angular separation is chosen due to the large angular uncertainty of LAT data at lower energy. IceCube neutrino angular uncertainty is about 1$^\circ$. A 10$^\circ$ angular separation between a photon-neutrino pair is an upper limit that guarantees the doublet searches. 
We also apply a temporal cut ($\Delta$T) of $\pm$50 s between neutrino and photon arrival times.  

\section{Statistical Methods}
\label{stat}

\subsection{Likelihood}
To perform the archival coincidence analysis between IC40/59 and Fermi LAT data, an unbinned log-likelihood function ($\lambda$) is used. This metric depends on event arrival direction and its uncertainty as well as a background rejection term and is defined as follows:
\begin{equation}
\label{llh}
\lambda = 2\ln\left(P_{\rm{LAT}}(\vec{x} | \hat{x}_\gamma) P_{\rm{IC}}(\vec{x} | \hat{x}_\nu)\right) - 2\ln\left(B(\hat{x}_\gamma) \right),
\end{equation}

\noindent where $\lambda$ is the likelihood parameter, $\hat{x}_\gamma$ and $\hat{x}_\nu$ are the arrival direction of the photon and neutrino events, respectively, $\vec{x}$ is the best fit position, $B(\hat{x}_\gamma)$ is the Fermi LAT background rejection term that will be briefly discussed in Section~\ref{bkgd}, $P_{\rm{LAT}}(\vec{x} | \hat{x}_\gamma)$ and $P_{\rm{IC}}(\vec{x} | \hat{x}_\nu)$ are the energy dependent point spread functions (PSF) of the LAT and IceCube observatories, respectively. In this metric, large values of $\lambda$ indicate a higher probability correlated pair.

The PSF for IceCube is given by a Gaussian function with an energy dependent spread~\cite{ICPSF}. The PSF for Fermi LAT is given by a two parameter King function~\cite{LATPSF}, deriving its energy dependency from empirically determined values. The best fit position of any doublet, $\vec{x}$, can be obtained analytically by solving a cubic equation.

\subsection{Background Rejection}
\label{bkgd}
%
%\begin{equation}
%    B(\vec{x}) = \int \Phi(\vec{x}, E) A(\vec{x}, E) dE \frac{\int \left(\frac{dN}{dE}\right)_\mathrm{test} dE}{\int \left(\frac{dN}{dE}\right)_\mathrm{test} A(\vec{x}, E) dE} \varpropto \frac{\mathrm{event}\; \mathrm{rate}(\vec{x})}{\mathrm{exposure}(\vec{x})},
%\end{equation}
%
%\noindent where $A(\vec{x}, E)$ is the exposure, $\Phi(\vec{x},E)$ is the photon flux, and $\left(\frac{dN}{dE}\right)_\mathrm{test}$ is a test spectral distribution taken to be $\propto E^{-2.41}$ which is the off-galactic diffuse background spectrum for the Fermi LAT observatory. 
The Fermi LAT background rejection term is a distribution that is proportional to the event rate over the exposure. For photons that end up in the galactic plane, the value of $B(\vec{x})$ is large, and therefore the pair will be given a less significant weight. A HEALPix map~\cite{healpix} with nside=64 (mean spacing of 0.91$^\circ$) is used to construct the background rejection map. 

\subsection{Null and Signal Distributions}
We generate a series of 10,000 scrambled datasets to estimate the null distribution. We scramble time and right ascension of each neutrino event while keeping declination and intrinsic neutrino properties (such as energy and directional resolution) unchanged. Time is shuffled using a uniform random number from a total time window of overlapping neutrino and photon datasets. 
We note that since LAT's motion is complex, we do not time-scramble the photon events.
Applying the spatial and temporal cuts on all of the IC40 and Fermi LAT neutrino-photon ($\nu-\gamma$) pairs using the method discussed in Section~\ref{data}, on average 2207$\pm 40$ pairs in each of the 10,000 null datasets are left.
Then a log-likelihood ($\lambda$) value is calculated for each $\nu-\gamma$ pair from the Equation~\ref{llh}.

Additionally, a series of 10,000 signal tests are created by injecting a limited number of forced coincidences into the null data.
Each signal dataset is obtained by matching a LAT photon event with each of the IceCube neutrino events. 
First, a random uniformly distributed source position is chosen on the sky as the true event position around which the messenger events will be distributed.
Then, the neutrino and photon events are randomly given a position on the sky centered around the north pole being the true event position.
This way, $\phi$ is uniformly distributed and $\theta$ is drawn from the PSF of each messenger.
These random positions are then rotated into space by transforming the north pole into the randomly chosen true event position. 
The photon energy is drawn from the convolution between the normalized power law distribution with an arbitrary spectral index of 2.2 and the LAT exposure function. 
Temporal coincidence is ensured through small time stamp offset. 
The angular separation between each photon and each neutrino is calculated and only pairs within the cut of $10^\circ$ are accepted and then $\lambda$ is calculated for each pair. 

\begin{figure}[!hb]
\begin{center}
\includegraphics [trim=0cm 0cm 0cm 0cm,scale=0.47]{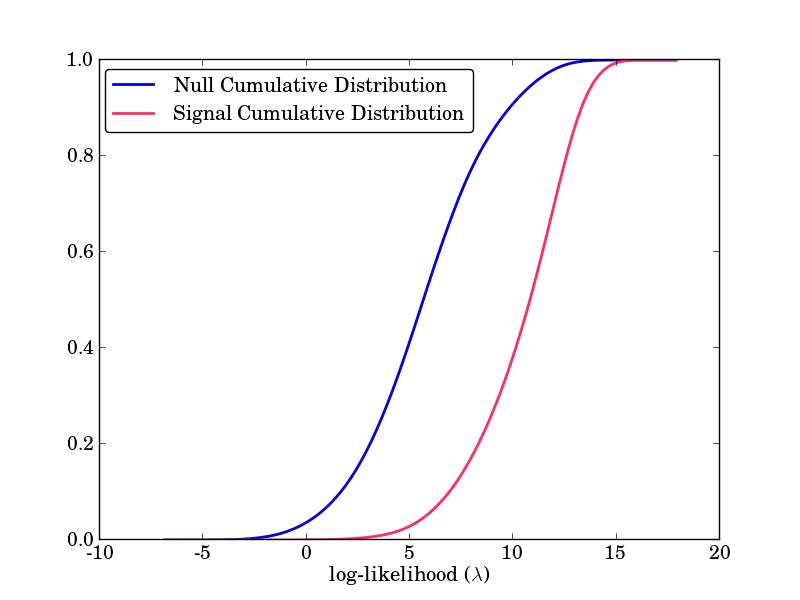}
\caption{Histograms of log-likelihood ($\lambda$) values of the null and signal distributions. Both distributions are normalized and cumulative. Signal distribution shows higher $\lambda$ values, as expected.}
\label{nullsigdist}
\end{center}
\end{figure}

Figure~\ref{nullsigdist} shows the normalized cumulative histograms of $\lambda$ values for both null and signal distributions using the described 10,000 datasets for IC40 and Fermi LAT coincidences. 

\subsection{Defining Statistical Excess}
\begin{figure}[!hb]
\begin{center}
\includegraphics [trim=0cm 0cm 0cm 0cm,scale=0.47]{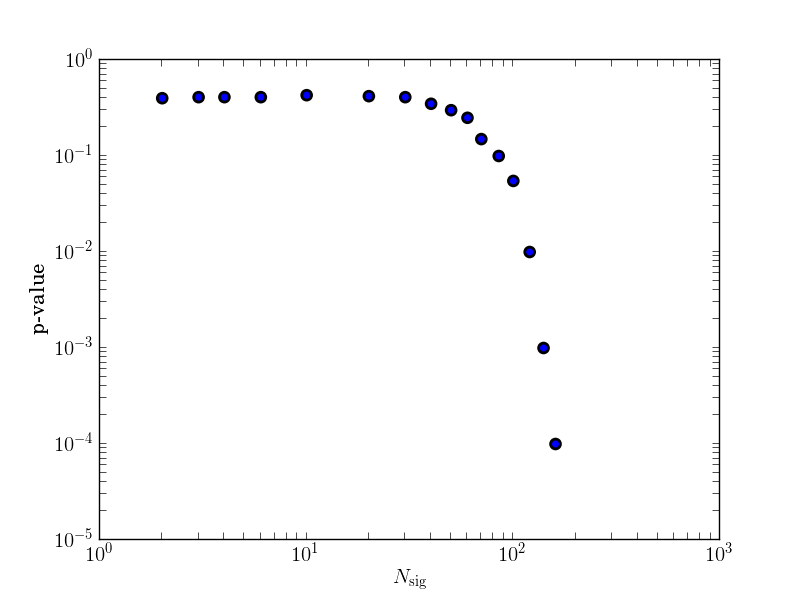}
\caption{Anderson Darling (AD) test statistic p-value versus number of injected signal photons ($N_{\rm{sig}}$). This plot helps us to obtain an estimate for the analysis effectiveness. The p-value obtained from the AD test suggests about 70-100 signal events.}
\label{pvalue}
\end{center}
\end{figure}

We perform an Anderson-Darling (AD) test~\cite{AD} on the null and signal likelihood distributions to determine if there is a statistical excess. 
%The AD test statistic is defined as follows:
%\begin{equation}
%    \kappa_\mathrm{AD} = n \int_{-\infty}^\infty{\frac{\left(F_n(x) - F(x)\right)^2}{F(x)\left(1 - F(x)\right)}d F(x)},
%\end{equation}
%\noindent where $F_n(x)$ is the cumulative distribution that we are testing, and $F(x)$ is the distribution that we are testing against. 
A p-value can be obtained from this test statistic using an AD k-sample test~\cite{ksample}. 
We inject signal photons randomly chosen from a single signal run with spectral index of 2.2. 
The results of obtained AD test statistic p-value is plotted versus the number of injected signal photons ($N_{\rm{sig}}$) in Figure~\ref{pvalue}. 
The injected signals are added to the null datasets and the new signal datasets is tested against the null hypothesis. 
This plot shows the ability of this method to distinguish injected signal from background. 
In other words, it helps us to obtain an estimate for the analysis sensitivity. 
We conclude from this plot that our analysis is sensitive to presence of $<100$ $\nu-\gamma$ pairs in full dataset. 

\section{Results}
\label{results}
\subsection{Unblinding Data}
\begin{figure}[!ht]
\begin{center}
\includegraphics [trim=0cm 0cm 0cm 0cm,scale=0.47]{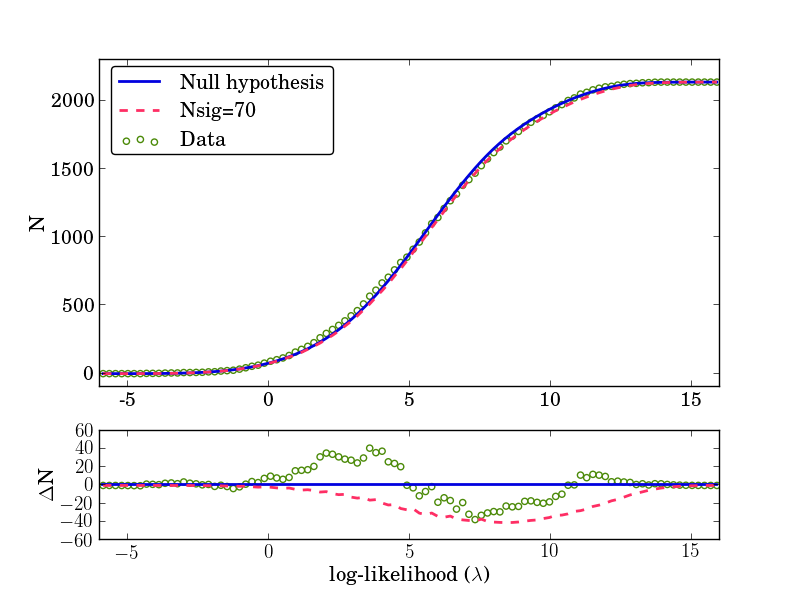}
\caption{Top: cumulative distribution of log-likelihood values of the IC40 and Fermi LAT coincidences after unblinding the data. A theoretical signal model with 70 injected signals ($N_{\rm {sig}}$=70) and the null hypothesis are plotted for comparison. All distributions are normalized and multiplied by 2138, which is the total number of coincidences after unblinding the data. Bottom: The residuals of data and $N_{\rm {sig}}$=70 versus the null hypothesis.}
\label{unblinded}
\end{center}
\end{figure}

Here we only present the results of unblinding the IC40 and Fermi LAT data. 
Unblinding the IC40 data and performing the coincidence analysis with Fermi LAT data result in 2138 coincidences applying the cuts of $\Delta\theta<10^\circ$ and $|\Delta \rm{t}|<50$ s. The AD test statistic on this dataset provides a p-value of about 4\%. 
This p-value is consistent with simulations including the injection of 70-100 signal events. 
The $\lambda$ distribution plot for unblinded data is presented in Figure~\ref{unblinded}. 
%The p-value obtained from the AD test suggests about 70-100 signal events within the dataset so a theoretical signal distribution with 70 injected signal events is also plotted to further compare the results. 
%, although the cumulative distributions (Figure~\ref{unblinded}) do not match particularly well. 
The null hypothesis and a theoretical signal model with 70 injected signals ($N_{\rm {sig}}$=70) are also plotted for comparison. 
Note that all distributions are normalized and multiplied by 2138, which is the total number of coincidences after the unblinding. 
The bottom plot in Figure~\ref{unblinded} shows the residuals of data and $N_{\rm {sig}}$=70 versus the null hypothesis to better represent the differences between the distributions. 
Adding 70 injected $\nu-\gamma$ pairs to the null data increases the number of high-$\lambda$ events. 
This explains the decrease observed in the residual plot of the two normalized cumulative distributions ($N_{\rm {sig}}$=70 and the null hypothesis) shown in red dashed line.
Although the cumulative distribution is not a particularly good match to the simulated distribution for $N_{\rm {sig}}$=70, we nonetheless take this as our model in defining a future three vetting tests for the data.
 %however we need to further vet this result using various tests. 

\subsection{Vetting the Signal}
Three different further tests are performed to explore whether real cosmic $\nu-\gamma$ pairs are present in the data at $N_{\rm {sig}}$=70 level. 
We perform the tests only on events with high-$\lambda$ values which are more likely to be signal.
To find the threshold above where we consider the $\nu-\gamma$ pairs ($\lambda_{\rm{cut}}$), we calculate the signal to noise ratio (SNR) for different $\lambda_{\rm{cut}}$ values. SNR is defined by the number of injected signals with $\lambda>\lambda_{\rm{cut}}$ divided by the square root of total number of pairs (injected signal + null data) also with $\lambda>\lambda_{\rm{cut}}$.
We find the maximum SNR at $\lambda_{\rm{cut}}$=11. 
Only events with $\lambda>\lambda_{\rm{cut}}$ are considered in the following tests. 

%\begin{figure}[!ht]
%\begin{center}
%\includegraphics [trim=0cm 0cm 0cm 0cm,scale=0.45]{snr-lambda.png}
%\caption{Signal to noise ratio (SNR) is plotted versus log-likelihood to find the $\lambda_{\rm{cut}}$.}
%\label{lambdacut}
%\end{center}
%\end{figure}

The first test is multiplicity, i.e. the number of photons in coincidence with a single neutrino.
The mean multiplicity of events with high-$\lambda$ is $\approx 2.17$ for real data whereas this number is $\approx 2.08$ for the null hypothesis with a standard deviation of 0.15 between the corresponding values of the 10,000 datasets. 
%Figure~\ref{multiplicity} shows the average multiplicity of events with low and high log-likelihood values, separately.
%The dashed line and the shaded band show the background expected value for average multiplicity and its uncertainty, respectively. 
Comparison between the mean multiplicity values of $\nu-\gamma$ pair coincidences in data and the null hypothesis indicates no significant signal excess above the background expectations with a p-value of 0.25. 

\begin{figure}[!ht]
\begin{center}
%\subfigure[Multiplicity]{
%\includegraphics [trim=1cm 0cm 1cm 0cm,scale=0.4]{multiplicity-lcut-data.png}
%\label{multiplicity}}
\subfigure[$\Delta\rm{t}$]{\includegraphics [trim=1cm 0cm 1cm 0cm,scale=0.32]{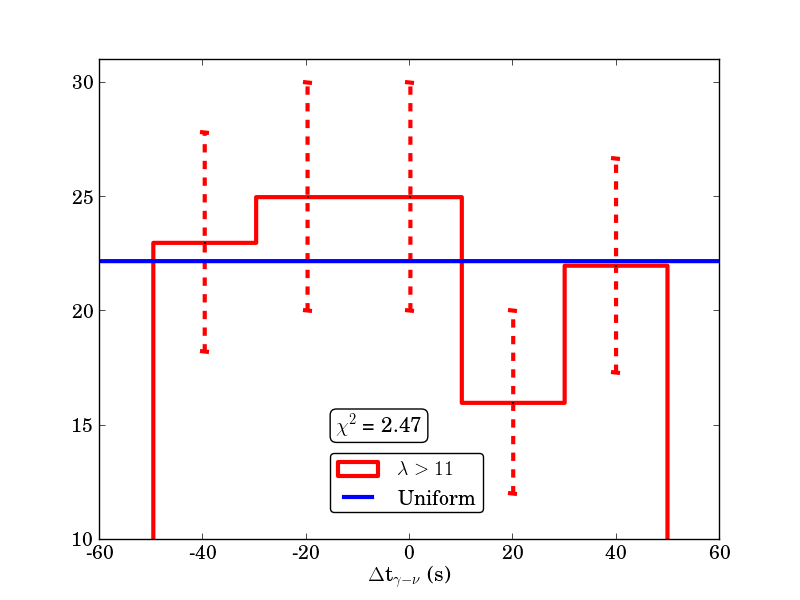}
\label{deltat}}
\subfigure[Clustering]{\includegraphics [trim=0cm 0cm 1cm 0cm,scale=0.41]{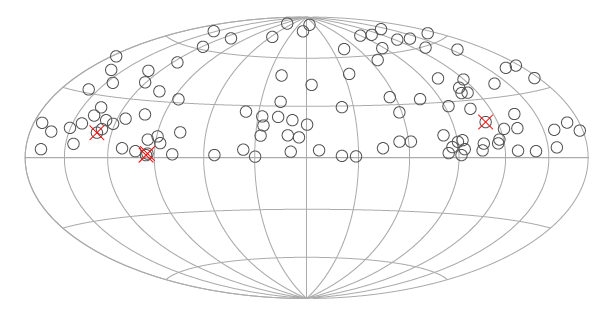}
\label{sourcemap}}
\caption{Vetting the signal: 
%(a) Multiplicity: the average event multiplicity is plotted for both low and high $\lambda$'s. The dashed horizontal line shows the average multiplicity of null data and the shaded region shows its uncertainty band. 
(a) $\Delta \rm{t}$: the histogram of the time difference between the neutrino and the photon in each pair is plotted and compared with a flat uniform distribution. 
(b) Clustering: source map of the high log-likelihood $\nu-\gamma$ pairs (shown by grey circles). In total, there are six $\nu-\gamma$ pairs that lie within $2^\circ$ of another pair, which is less than the result we get from null hypothesis. 
Red crosses are the locations of the four neutrinos that contribute to the total of six $\nu-\gamma$ pairs. 
These tests indicate no significant signal excess in IC40 and Fermi LAT datasets.}
\label{3tests}
\end{center}
\end{figure}

The next test is to look for a significant difference between different $\Delta\rm{t}$ bins. $\Delta\rm{t}$ is the time difference between the IC40 neutrino in coincidence with the Fermi LAT photon event. Figure~\ref{deltat} shows the $\Delta\rm{t}$ histogram for five time bins compared to a uniform distribution. The reduced chi-square value is approximately 0.5, which indicates consistency with a flat uniform distribution, further suggesting absence of signal.   

%\begin{figure}[!ht]
%\begin{center}
%\includegraphics [trim=0cm 0cm 0cm 0cm,scale=0.45]{deltat-hist-realdata-lambdacut-chisq.png}
%\caption{Time difference between IC40 neutrino in coincident with Fermi LAT photon event ($\Delta\rm{t}$) histogram}
%\label{deltat}
%\end{center}
%\end{figure}

Finally, we test clustering of the high log-likelihood $\nu-\gamma$ pairs which would reveal bright/repeating sources. Figure~\ref{sourcemap} shows the source map of these pairs. We see that only six $\nu-\gamma$ pairs lie within $2^\circ$ of another pair. 
In total, four neutrinos are contributed in creating these six pairs which are indicated by red crosses on the sky plot. 
Six $\nu-\gamma$ pairs are shown to be less than the average result of 12.9 such clustered $\nu-\gamma$ pairs we get from our 10,000 scrambled datasets; roughly no evidence for a contribution for cosmic $\nu-\gamma$ pairs.

%\begin{figure}[!ht]
%\begin{center}
%\includegraphics [trim=0cm 0cm 0cm 0cm,scale=0.4]{Source-map.png}
%\caption{Source map of the high log-likelihood neutrino-photon pairs.}
%\label{sourcemap}
%\end{center}
%\end{figure}

%Figure~\ref{clustered} shows the number of clustered neutrino-photon pairs from within the high log-likelihood subset. 

%\begin{figure}[!ht]
%\begin{center}
%\includegraphics [trim=0cm 0cm 0cm 0cm,scale=0.4]{culstered-neutrinos.png}
%\caption{Source map of the high log-likelihood neutrino-photon pairs.}
%\label{clustered}
%\end{center}
%\end{figure}

\section{Conclusions}
\label{conclusions}
We performed an archival analysis on neutrinos from IceCube observatory in coincidence with Fermi LAT photon events, both from public datasets. 
Several statistical tests on observed data using the background and signal datasets were conducted. 
The Anderson-Darling test statistic showed about 70 signal out of 2138 found coincidences in IC40-Fermi LAT analysis, however multiplicity, $\Delta\rm{T}$, and clustering tests showed no significant signal excess. 
The results of IC59 and Fermi LAT and the combined IC40/59 datasets and Fermi LAT will be presented in a future publication. \\

\noindent{\bf Acknowledgement.} The authors acknowledge support from the National Science Foundation under grant 003403953 and the Institute for Gravitation and the Cosmos of the Pennsylvania State University.
%This research is supported by grants from the U.S. National Science Foundation (003403953) and the Institute for Gravitation and the Cosmos of the Pennsylvania State University.


\begin{thebibliography}{99}
\bibitem{AMON2013} 
M. W. E. Smith et al., {\it The Astrophysical Multimessenger Observatory Network (AMON)}, Astroparticle Physics 45 (2013) 56 [{\tt arXiv:1211.5602v1}].

\bibitem{gociazi} G. Te\v{s}i\'{c} and A. Keivani, {\it AMON: Transition to Real-Time Operations}, in proceedings of {\it ICRC}, \pos{PoS(ICRC2015)329}, 2015.

\bibitem{wb}
E. Waxman, J. Bahcall, {\it High Energy Neutrinos from Cosmological Gamma-Ray Burst Fireballs}, Physical Review Letterts 78 (1997) 2292 [{\tt arXiv:astro-ph/9701231}].

\bibitem{rachen}
J. P. Rachen, P. M\'{e}sz\'{a}ros, {\it Photohadronic Neutrinos from Transients in Astrophysical Sources}, Physical Review D 58 (1998) 123005 [{\tt arXiv:astro-ph/9802280}].

\bibitem{IC3yrs}
IceCube Collaboration, M. G. Aartsen et al., {\it Observation of High-Energy Astrophysical Neutrinos in Three Years of IceCube Data}, Physical Review Letters 113 (2014) 101101 [{\tt arXiv:1405.5303}].

\bibitem{ICPSF}
IceCube Collaboration, R. Abbasi et al., {\it Time-Integrated Searches for Point-Like Sources of Neutrinos with 40-String IceCube Detector}, The Astrophysical Journal 732 (2011) 18 [{\tt arXiv:1012.2137}]. 

\bibitem{LATPSF}
Fermi-LAT Collaboration, M. Ackermann et al., {\it Determination of the Point-spread Function for the Fermi Large Area Telescope from On-orbit Data and Limits on Pair Halos of Active Galactic Nuclei}, The Astrophysical Journal 765 (2013) 54 [{\tt arXiv:1309.5416}].

\bibitem{healpix}
K. M. Gorski et al., {\it HEALPix: A Framework for High-Resolution Discretization and Fast Analysis of Data Distributed on the Sphere}, The Astrophysical Journal 622 (2005) 759 [{\tt arXiv:astro-ph/0409513}]. 

\bibitem{AD}
T. W. Anderson and A. D. Darling, {\it Asymptotic Theory of Certain "Goodness of Fit" Criteria based on Stochastic Processes}, The Annals of Mathematical Statistics 23 (1952) 193.

\bibitem{ksample}
F. W. Scholz and M. A. Stephens, {\it K-Sample Anderson-Darling Tests}, Journal of the American Statistical Association 82 (1987) 918.

\end{thebibliography}
\end{document}